\documentclass[journal]{IEEEtran}
\usepackage{enumitem}
\usepackage[english]{babel}
\usepackage{blindtext}
\usepackage{epsfig}
\usepackage{xcolor}
\usepackage{balance}
\usepackage{url}
\usepackage{comment}
\usepackage{lipsum}
\usepackage[hang,flushmargin]{footmisc}

\setlist[itemize]{leftmargin=*}

\begin{document}
	
\title{Internet of Everything (IoE) - \textit{From Molecules}\\ \textit{to the Universe}}
	
\author{Ozgur~B.~Akan,~\IEEEmembership{Fellow,~IEEE,}
		Ergin~Dinc,~\IEEEmembership{Member,~IEEE,}
		Murat~Kuscu,~\IEEEmembership{Member,~IEEE,}
		Oktay~Cetinkaya,~\IEEEmembership{Member,~IEEE,}
		Bilgesu~A.~Bilgin,~\IEEEmembership{Member,~IEEE}
		\thanks{\hspace{2mm} O. B. Akan, E. Dinc, and B. A. Bilgin are with the Internet of Everything (IoE) Group,	Department of Engineering, University of Cambridge, Cambridge CB3 0FA, UK (e-mail: \{oba21, ed502, bab46\}@cam.ac.uk).}
		\thanks{\hspace{2mm} M. Kuscu and O. Cetinkaya are with the Department of Electrical and Electronics Engineering, Koc University, Istanbul 34450, Turkey (e-mail: \{mkuscu, ocetinkaya\}@ku.edu.tr).}
		\thanks{\hspace{2mm} O. B. Akan is also with the Department of Electrical and Electronics Engineering, Koc University, Istanbul 34450, Turkey (e-mail: akan@ku.edu.tr).}
		\thanks{\hspace{2mm} This work was supported in part by the ERC project MINERVA 
		(ERC-2013-CoG \#616922), AXA Research Fund (AXA Chair for Internet of Everything at Koc University), The Scientific and Technological Research Council of Turkey (TUBITAK) under Grant \#120E301, and EU’s Horizon 2020 Research and Innovation Programme through the Marie Skło-dowska-Curie Individual Fellowship under Grant Agreement \#101028935.}}
	
\IEEEpeerreviewmaketitle
\maketitle
	
\begin{abstract}
The universe is a vast heterogeneous network of interconnected entities that continuously generate and exchange information through various forms of interactions, some of which are yet to be discovered. Internet of Everything (IoE) framework, inspired by the ubiquitous and adaptive connectivity and the seamless interoperability within this universal network, puts forward a new road map beyond the conventional Internet of Things (IoT) towards maximizing the resolution of our interface with the universe to enable unprecedented applications. The first pillar of this road map is to reveal novel and tangible interconnections between seemingly noninteracting branches of IoT, which we call IoXs with X referring to their application domains, e.g., Internet of Energy (IoEn), Internet of Vehicles (IoV). The second pillar is to develop new IoXs that can complement the existing ones to complete the overall IoE picture and match its networking traits to that of the universe for a seamless and all-embracing cyber-physical interface. The objective of this paper is to evaluate the potential of this holistic IoE approach to expand the limited application landscape of the current IoT practice on a scale ranging \textit{from molecules to the universe}. To this end, we identify several potential interaction pathways among IoXs and introduce novel and emerging IoXs that are essential to the comprehensiveness of IoE. We also discuss the potential applications that can be enabled by such interconnections within the IoE framework and identify the associated challenges. 
\end{abstract}
	
	
\section{Introduction}
	
Our accumulated scientific knowledge suggests that the~universe is a heterogeneous network of `everything',~ranging from molecules to the planets. Some of the most complex~phenomena, e.g., evolution and consciousness, are~believed to be rooted in complex interaction networks that~create more information than the interacting parts. This ubiquitous connectivity of the universe and the `more~than~the~sum'~characteristics of the underlying heterogeneous networks are~the two main traits inspiring the emerging Internet of Everything (IoE) approach. 
	
IoE is a big step forward beyond the conventional IoT, which has long been under the scope of both academia and industry, with several applications, e.g., smart meters in energy grids, industrial and agricultural wireless sensor networks (WSNs), that found their way into the market. One of the main challenges of IoT is the lack of interaction between its branches, i.e., Internet of Xs (IoXs), each targeting only a specific application domain (X). For example, the Internet of Vehicles (IoV) aims establishing networks of smart vehicles to optimize traffic flow at a lower environmental/operational cost. However, it has no direct liaison with other domains, such as industrial plants or agricultural fields, which could benefit from that networking approach of IoV to attain a better efficiency vs cost index. This apparent disconnection between IoXs leads to a short-sighted perspective missing out on many opportunities that lay in the interaction of heterogeneous technologies, which can generate higher value than individual IoXs. 
	
IoE takes a holistic approach and aims to unify the existing IoXs based on novel interaction pathways defined between them. Such interactions include a blockchain-based energy market enabling consumers to trade energy directly with each other and with the grid (instead of retailers) via energy tokens, which is a by-product of the Internet of Money (IoM) and the Internet of Energy (IoEn) merger. IoV can be combined with those two to achieve peer-to-peer (P2P) vehicle charging (using the same tokens), minimizing the waiting times and pressure on scarce resources through efficient energy cooperation between peers. This ambitious goal, however, requires the optimal match of donor and recipient vehicles by tracking their locations and battery levels in real-time. That can be met by the Internet of Space (IoSP) seamlessly communicating with the IoM, IoEn, and IoV within the IoE framework. 
	
\begin{figure*}[!t]
	\centering
	\includegraphics[width=0.87\textwidth]{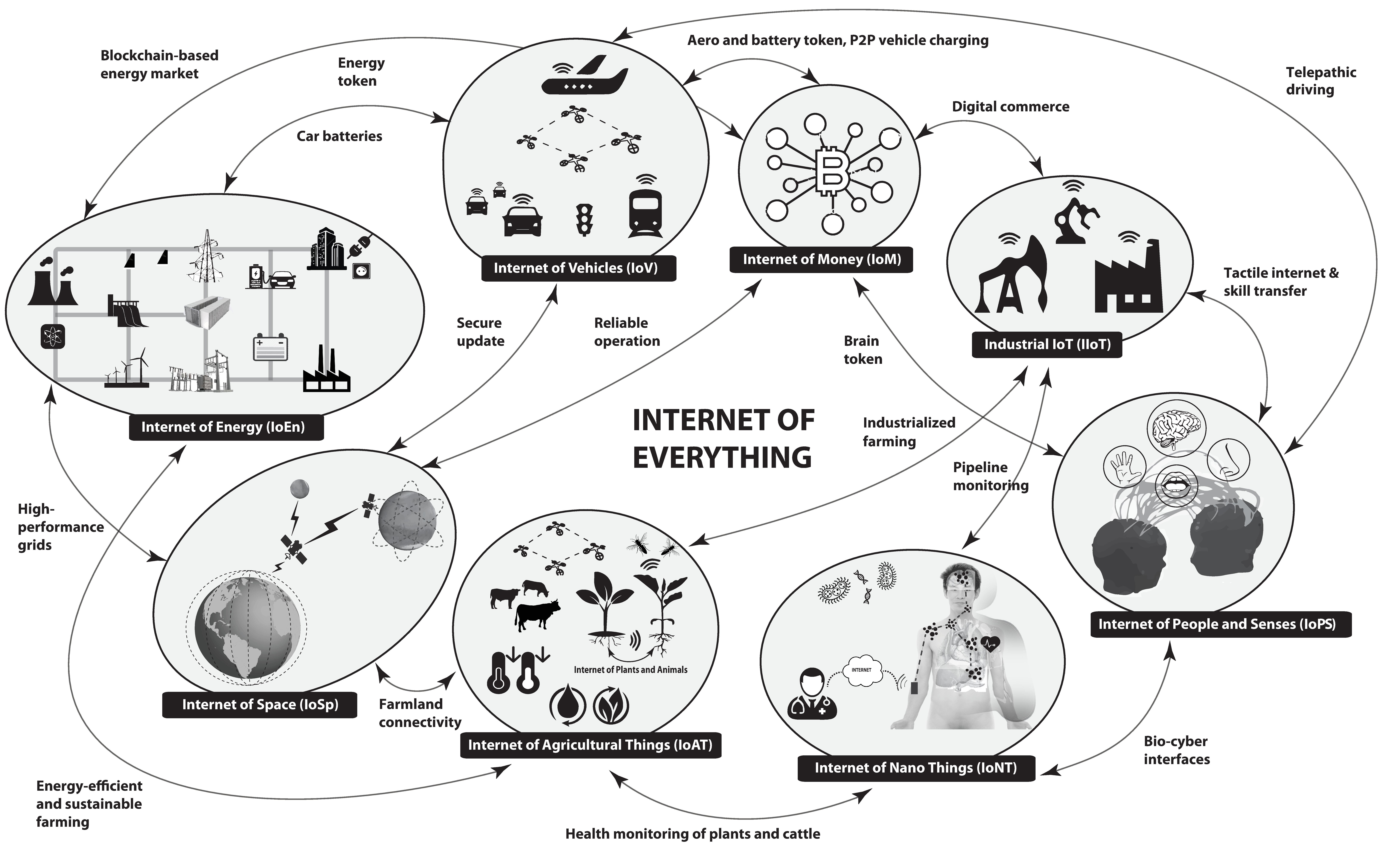}
	\caption{The upcoming IoE landscape with its major components -IoXs.}
	\label{fig:Fig1}
\end{figure*}
	
To match the ubiquitous connectivity and heterogeneous networking characteristics of the universe, IoE also integrates new IoXs into its framework. Internet of Nano Things (IoNT), for example, is poised to increase the resolution of cyber-physical interfaces and bring connectivity into uncharted territories, e.g., inside the human body, with the networks of smart biological agents. Internet of People and Senses (IoPS), as another example, refers to the conceptual transfer of information and even skills between humans besides the nonverbal communication of senses, e.g., olfaction and gustation. These will ultimately enable a seamless cyber-physical interface with a high spatiotemporal resolution and create unprecedented opportunities to monitor and control the natural interaction pathways and develop novel applications. That, of course, requires overcoming many challenges, such as interoperability, miniaturization, and energy efficiency.
	
\newpage
Our objective in this paper is to bring the upcoming IoE revolution to attention. Hence, we first discuss the state-of-the-art in key IoXs (Fig.~\ref{fig:Fig1}), which are the essential components of the IoE framework, such as the Industrial Internet of Things (IIoT), Internet of Agricultural Things (IoAT), IoM, IoV, and IoEn. We also introduce and discuss emerging IoXs, such as IoNT, IoPS, and Internet of Space (IoSp), which complement the existing ones to complete the overall IoE picture. We identify the opportunities and the challenges in advancing individual IoXs and creating interconnections between them within the IoE framework. Lastly, we provide a road map for the evolution of the IoE framework within the next 30 years. 
	
\section{Internet of Xs (IoXs)}
	
The introduced IoE vision consists in the seamless interaction of heterogeneous technologies and applications integrating all the essential elements of the universe, including inanimate and living entities. This section introduces and reviews these major technologies as the building blocks of the IoE, i.e., IoXs, which are categorized based on their application areas spanning the whole universe, starting from the molecular scale. 
	
\begin{figure*}[!ht]
	\centering
	\includegraphics[width=0.87\textwidth]{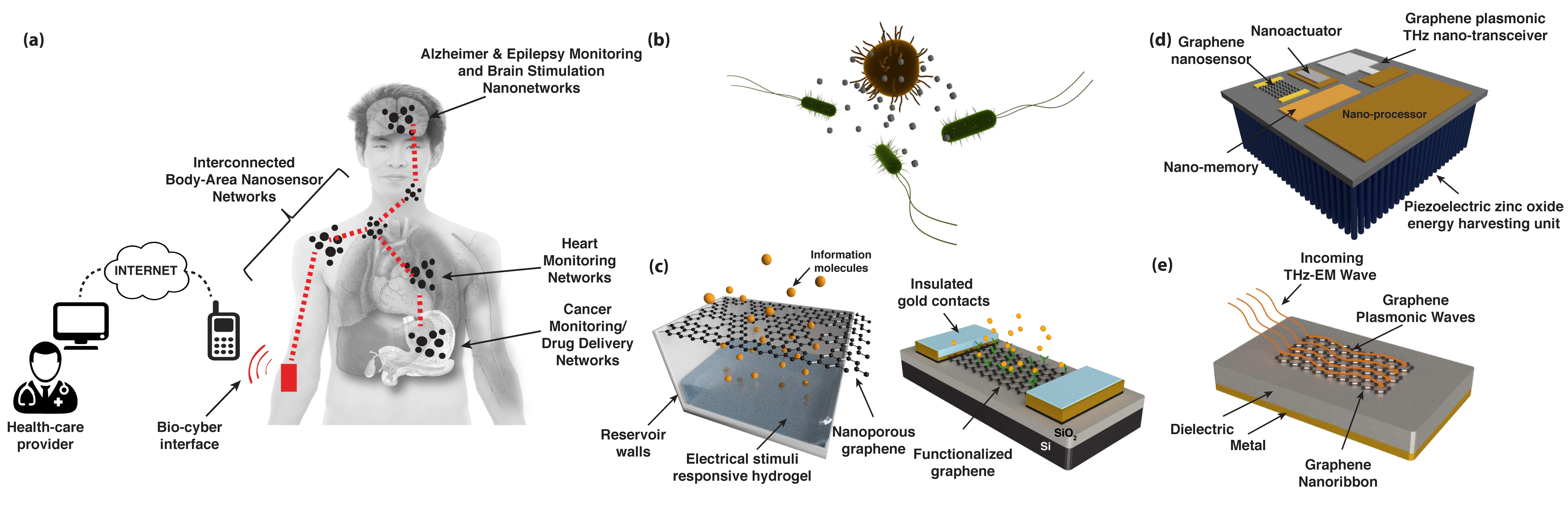}
	\caption{(a) Conceptual drawing of a continuous health monitoring application of IoNT. (b) MC among engineered bacteria within IoNT. (c) Graphene-based nanoscale MC transmitter \& receiver architectures. (d) EH nanomachine architecture for IoNT. (e) Graphene plasmonic nanoscale THz transceiver~architecture.}
	\label{fig:Fig2}
\end{figure*}
	
\subsection{Internet of Nano Things (IoNT)}
Nanotechnology has enabled the manipulation of individual atoms to develop new nanomaterials with exceptional characteristics and the design of nanoscale machines interfacing with the physical universe at the atomic level. The idea of IoNT, as illustrated in Fig. \ref{fig:Fig2}, lies in interconnecting nanomachines of different functionalities to overcome their resource limitations and increase their operational capabilities, besides integrating these nanonetworks with the conventional electromagnetic (EM) wireless networks through nano-macro and bio-cyber interfaces to enable unprecedented applications, such as intrabody continuous health monitoring \cite{akyildiz2010internet}. IoNT will be the most abundant component of the IoE in terms of the number of connected nodes and the amount of data generated.
	
Research in this field has been focusing on physical layer design, where terahertz (THz)-band EM \cite{lemic2021survey} and molecular communications (MC) \cite{akan2017fundamentals} are the most promising approaches to enable reliable information transfer at the nanoscale. MC, being already realized by living cells, provides a more biocompatible ground for developing artificial nanonetworks; thus, it has attracted the most attention. IoNT research prioritizes developing channel models, nano-transceiver architectures, modulation/detection techniques, and communication protocols for MC \cite{kuscu2019transmitter}. However, there is still an immense discrepancy between the complexity of the developed methods and the resource limitations of nanomachines. Open research challenges for IoNT are summarized as follows \cite{10.3389/frcmn.2021.733664}:
\begin{itemize}
	\item \textbf{Miniaturization:} IoNT requires pushing the size of network nodes down to nanoscale and devising communication methods compatible with these miniature devices to enable true cyber-physical interfacing with high spatiotemporal resolution. However, no IoNT device implementation has so far achieved all IoNT functionalities, e.g., nanoscale communication, interface with macroscale networks, and harvest energy. The emergence of novel nanomaterials, e.g., graphene, with extraordinary optoelectronic and chemical properties, is promising for developing novel IoNT interfacing and communication modalities, e.g., the use of plasmons and molecules for sensing and information exchange.
	\item \textbf{Ubiquitous~Connectivity~and~Interoperability:}~Envisioned IoNT applications span various harsh environments, e.g., intrabody, imposing several connectivity~challenges, which cannot be overcome by conventional communication methods, thus necessitates novel bio-inspired techniques like MC. Developing bio-cyber interfaces to connect nanonetworks, including those composed of bacteria- and nanomaterial-based networks, with each other and to the Internet is another challenge towards interoperability, requiring fundamental research on exploiting novel nanomaterials to create seamless interfaces and transceivers accommodating both molecular and EM communication modalities. 
	\item \textbf{Self-sufficiency:} Conventional means of energy supply~and storage are not feasible at nanoscale since miniaturization introduces strict limitations, especially for storage. This challenge can be tackled by developing more bio-inspired techniques (e.g., energy-efficient MC), novel energy harvesting (EH) methods (e.g., intrabody chemical EH from glucose), low-complex communication protocols, and competent molecular or wireless power transfer (WPT) techniques. 
	\item \textbf{Big Data:} Envisioned IoNT applications require developing novel data analytics tools to exploit the unprecedentedly big data generated by billions of densely deployed nanomachines, which consist of mostly unstructured data because of the limited computation capabilities of the nanomachines. 
\end{itemize} 
	
The interconnection of IoNT with other IoXs can introduce an infinite variety of new directions to the IoE realm. For example, pipeline monitoring in industrial plants (w/IIoT) and oil\&gas distribution systems (w/IoEn) with nanoscale sensors detecting corrosion, leaks, blockages, or impurities can satisfy regulatory requirements. Similarly, networked nanorobots injected into the circulatory system of humans can timely diagnose any disease or implication, e.g., blood clots, tumors, and treat/remove them with dedicated drug delivery or actuating capabilities. IoNT can also cooperate with IoPS towards achieving bio-cyber interfaces, translating biochemical signals delivered by intra-body nanonetworks into electrical terms, and vice versa, for seamless biotic-abiotic interactions.
	
\begin{figure*}[!t]
	\centering
	\includegraphics[width=0.87\textwidth]{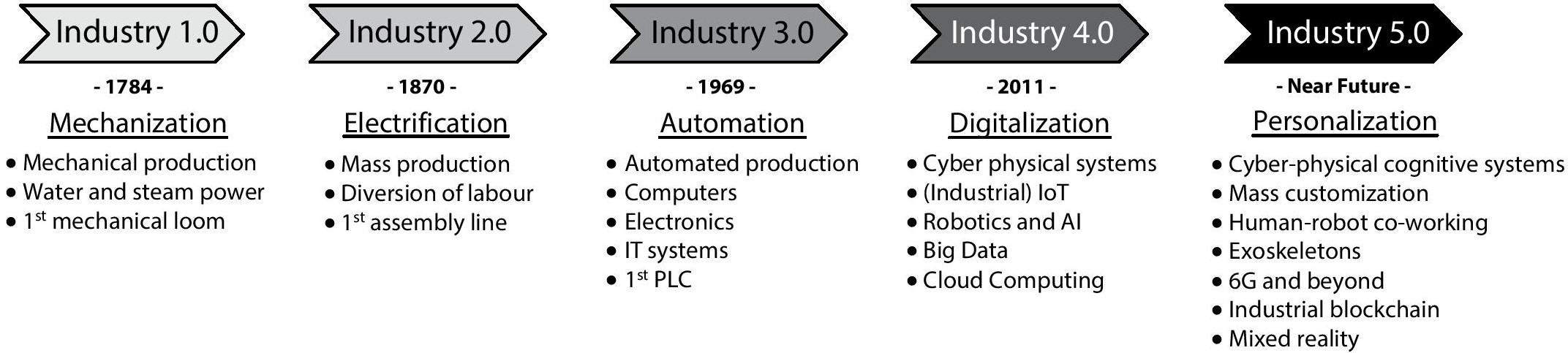}
	\caption{History of the industrial revolution, revealing how the adaptation of the Internet marked the new epoch (Industry 4.0) almost a half-century earlier.}
	\label{fig:Fig3}
\end{figure*}
	
\subsection{Internet of People and Senses (IoPS)}
	
Sharing human cognitive functionalities and senses through the Internet, i.e., IoPS, can lead to the most groundbreaking applications of the IoE framework. The interconnection of people's brains, i.e., Brainets, for an advanced network-scale consciousness leading to higher-level intelligence and for new forms of direct conceptual communication and collaboration among people, is the ultimate goal of the IoPS vision \cite{jiang2019brainnet}. However, IoPS adopts many other technologies gaining maturity. For example, the Tactile Internet, i.e., real-time sharing of touch and actuation, enabling the transfer of skills and labour, has already found applications in remote healthcare, education, and gaming. Digital communication of smell and taste is also gaining momentum, promising new forms of social networking based on non-verbal communication modalities~\cite{calvi2020scent}. Yet, the IoPS faces fundamental research challenges explained below: 
\begin{itemize}[leftmargin=*]
	\item \textbf{Miniaturization:} Similar to the IoNT, IoPS calls for~substantial efforts to develop miniaturized and biocompatible bio-cyber/neural interfaces that are capable of transducing sense patterns and cognitive process outcomes into digital signals and recreating them at the receiving end. In this direction, the emergence of nanomaterials and the IoNT technology are promising for developing interfaces with living cells at very high spatiotemporal~resolutions.
	\item \textbf{Big Data:} Exploiting the big data generated by nanoscale bio-cyber/neural interfaces requires a comprehensive understanding of the complex human brain and isolating~useful patterns therein. This challenge has been targeted~by~the Human Brain Project\footnote{The Human Brain Project. Source: https://www.humanbrainproject.eu/en/} and the Brain Initiative\footnote{The Brain Initiative. Source: https://braininitiative.nih.gov/}, the two major research projects respectively supported by the EU and US, ominating further efforts in the area. 
\end{itemize} 
	
Under the IoPS vision, we can connect ourselves to the Internet, not only for health monitoring, etc., but also for gaining new capabilities, e.g., mind control over electronic devices, Internet access by thought, and sensing in a wider EM and acoustic spectrum. Moreover, our bodily functions, such as body heat/fluids and brain activities, can be used to validate blockchain transactions and thus mine cryptocurrencies or \textit{brain tokens} (w/IoM). Last but not least, we can bring telepathic driving (w/IoV) into existence via brain-to-vehicle technology, revolutionizing the autonomous car industry with mind-controlled steering, powertrain, and brake systems.
	
\subsection{Internet of Industrial Things (IIoT)}
	
The fourth industrial revolution, namely Industry 4.0 \cite{frank2019industry}, strives for the combination of Internet and future-oriented technologies with already electrified and automated industrial machinery (Fig.~\ref{fig:Fig3}). It aims to improve industry services via digitizing the manufacturing process involved by means of increasing the effectiveness of collaboration between machines and by improving product and service quality. To achieve this, the whole connected system of physical entities, e.g., factory equipment and products together with cyber-entities, collection of software performing optimal control of the physical processes, referred to as Cyber-Physical Systems (CPSs), are deployed into factories/industrial plants within the IoE context. Furthermore, the issues like safety, security, and surveillance at those places are significantly improved via WSNs. The forthcoming Industry 5.0 Era is expected to better fit into this purpose as well as the IoE framework by interconnecting several IoXs. Although the industry is among the first adopting the IoE, many challenges hinder widespread IIot utilization:
\begin{itemize}
	\item \textbf{Interoperability:} The integration of heterogeneous devices comprising the CPSs to be used, which vary with the type of industry and the processes involved in manufacturing, is a significant challenge. Since the efficacy and efficiency of the deployed CPSs heavily rely on the seamless cooperation of involved devices, problems in the interoperability of these devices directly translate into financial consequences.
	\item \textbf{Big Data:} Product line and quality optimization of manufacturing require big data analytics to be performed on CPSs, which are very case-sensitive; hence, require~customized solutions for each industry and even for each~workplace.
	\item \textbf{Security and Privacy:} Automation of product lines come with increased safety risks as the CPSs are prone to malfunctions and cyber attack-driven failures. Furthermore, the surveillance required for process control has the potential pitfall of restricting the privacy of workers at the workplace.
\end{itemize} 
	
IIoT is closely associated with many IoXs. For example, the digitalization of farming within the Industry 4.0 era enabled precision agriculture with advanced industrialized farming tools. Today, the interaction between IIoT and IoAT is moving towards an integrated system of systems solution through the seamless cooperation of weather data, farm equipment, and irrigation systems. IIoT in energy sectors (IoEn), as another combination, can minimize downtimes, balance supply\&demand, and achieve predictive maintenance. Similarly, the merger of IIoT with digital commerce (IoM) can avoid disruptions in the supply chain through data-driven insights, besides keeping better inventory and maintaining quality, referring to ever-efficient asset optimization and tracking.
	
\subsection{Internet of Vehicles (IoV)}
	
The number of devices connected to the Internet is projected to reach 41 billion by 2027\footnote{The Internet of Things 2020 Report. Source: \url{https://www.businessinsider.com/internet-of-things-report?r=US&IR=T}.}, of which~vehicles~represent~a~substantial portion. The integration of vehicles in the IoT domain and their interaction with other vehicles, pedestrians, (network) infrastructure, and road-side units, also termed vehicle-to-everything (V2X) communications, has converted the old Vehicular Ad Hoc Networks (VANETs) into the IoV concept \cite{ji2020survey}. The ultimate goal of IoV is to establish~a network of ``smart" vehicles that exchange information for coordinating their automated behaviour to minimize risks and maximize traffic flow at lower emission, cost, and energy consumption. 
	
IoV gathers specific hardware, software, network technologies, and third-party services to offer novel safety, mobility, and infotainment applications, which require~reliable communication between vehicles and their surroundings. Since cities become more interconnected and intelligent as days pass, they offer the ideal conditions for IoV proliferation while helping connected vehicles gradually transform into autonomous entities. However, several issues need addressing before IoV reaches its full potential:
\begin{itemize}
	\item \textbf{Security:} IoV integrates various technologies, standards, and services for the flawless operation of its components, which makes it vulnerable to malicious acts. As the IoV allows remote access to in-vehicle sensors, GPS, brakes, etc., successful attacks may result in serious casualties.
	\item \textbf{Reliability:} Reliable connection and perpetual connectivity are of utmost importance for the IoV, and network failures, malevolent attacks, and other bottlenecks can severely impact the whole infrastructure. Thus, the highly mobile and dynamic nature of the IoV necessitates spatial and temporal endurance to the changes in external factors, such as speed, location, and attackers, to ensure that the IoV components communicate without interruptions.
	\item \textbf{Big Data:} Currently, connected vehicles process about 1GB of data per second, but that number is expected to grow as more infrastructure goes online and becomes interconnected. Insufficient storage capacity or network latency can hamper cloud computing and damage the systems.
\end{itemize}
	
One of many novel applications emerging from IoV and IoX cooperation is P2P charge trading between electric vehicles (w/IoM \& IoEn). Safe and accurate vehicle-assistance systems, as another example, are achieved by networked miniature sensors (w/IoNT) spreading into every part of a vehicle to track proximity and environmental conditions in real-time. The farming industry also benefits from the IoV, with drones creating phenotypes by determining growth/health status through dynamic aerial monitoring of plants (w/IoAT). IoV's intersection with IoPS has introduced Social IoV, which gathers common interest groups, i.e., drivers, passengers, and transport authorities, to exchange traffic information for a better driving experience.
	
\subsection{Internet of Money (IoM) }
	
Cryptocurrencies have emerged to tackle the inefficiencies of the conventional banking system, e.g., long account opening and transaction processing times. They utilize blockchain technology \cite{dai2019blockchain}, which is based on a distributed ledger system with no central ledger, i.e., a bank approving transactions. Here, the cryptocurrency miners serve as the distributed ledger to generate cryptocurrencies, which offers improved reliability, flexibility, and security. Cryptocurrency wallets can be obtained immediately anywhere by anyone, and transactions can be made online in minutes without any border or cost.~Hence, cryptocurrencies are regarded as the money of the Internet,~i.e., IoM, and their value is stored by the connectivity of distributed ledger. Although cryptocurrency market capitalization once reached almost \$$3$T, IoM faces significant challenges: 
\begin{itemize}
	\item \textbf{Scarcity of Resources:} The number of cryptocurrency transactions is significantly increasing, and that causes an increase in the approval times. This issue can be tackled by increasing network resources, e.g., the number of cryptocurrency miners, subject to ongoing GPU shortage.
	\item \textbf{Energy Consumption:} Increasing network resources leads to higher energy consumption. According to the University of Cambridge researchers, bitcoin mining alone consumes $97.3$TWh of energy annually -nearly as much as Pakistan- as of Q3 2022\footnote{Cambridge Bitcoin Energy Consumption Index. Source: https://cbeci.org/}. Developing energy-efficient integrated circuits and algorithms can overcome this.
	\item \textbf{Standardization:} Since governments perceive cryptocurrencies as an enabler for illegal money traffic, there is no standardization for cryptocurrencies yet.
\end{itemize}
	
The distributed ledger system, underpinned by IoE devices, can offer many brand new applications. For example, a blockchain-based energy market can enable consumers to trade energy directly from the grid (instead of retailers) in a more secure way via energy tokens, a by-product of IoM and IoEn merger. IoV can be combined with those two to achieve P2P vehicle charging via battery tokens. Speaking of tokens, brain tokens can be used as a denomination for skill transfer between humans (w/IoPS), as mentioned earlier. Lastly, blockchain technology can assist better tracking of manufactured products and avoid disruptions in the supply chain (w/IIoT), thereby maximizing efficiencies.
	
\subsection{Internet of Energy (IoEn)}
	
\begin{figure}[!t]
	\centering
	\includegraphics[width=0.435\textwidth]{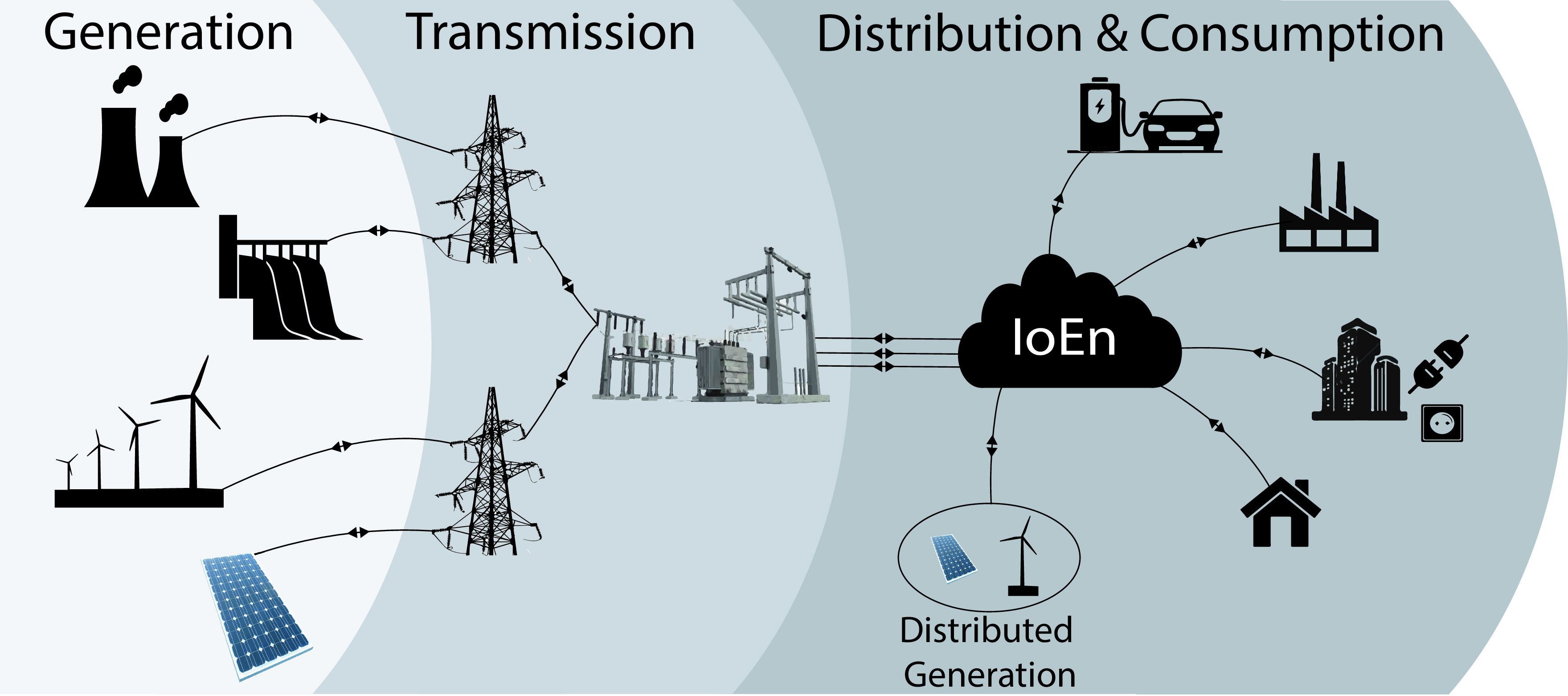}
	\caption{Smart and intelligent power grid in the IoEn framework.}
	\label{fig:Fig4}
\end{figure}

IoEn defines the modernization of energy production, transmission, distribution, and consumption by upgrading the existing energy infrastructure via the IoE. The current setting adopts a centralized approach, causing vast amounts of waste during transmission or production due to supply\&demand mismatch. Furthermore, it is not flexible to accommodate renewable energy plants at any scale. Thus, the Smart Grid concept is proposed to transform the existing grid into a decentralized and intelligent one using smart meters, actuators, and WSNs \cite{wang2017survey}. This concept automates grids through real-time monitoring of supply\&demand, as depicted in Fig.~\ref{fig:Fig4}, achieving higher efficiencies.
	
IoEn also covers powering WSNs, essential for IoE realization, where the limited battery life is a key issue. Non-deterministic battery depletion threatens sensor reliability, replacement of which means high maintenance costs and frequent disruptions. EH, together with WPT, can mitigate this by making IoE devices self-sufficient, i.e., energy-autonomous. However, the availability of harvestable sources, e.g., solar, often depends on external factors, imposing another challenge. One way to tackle that is to adopt multiple EH mechanisms, enabling higher reliability and self-sufficiency of sensors \cite{akan17internet}. Yet, energy-efficient sensing, computation and communication techniques should also be considered for uninterrupted operations in the~IoE. The other challenges of IoEn are as~follows:
\begin{itemize}
	\item \textbf{Interoperability:} Integrating small- and large-scale supply\&demand from various sources into decentralized and deregulated energy production increases the complexity of power grids. Thus, interoperability of energy systems and their monitoring via the WSNs is an important issue.
	\item \textbf{Privacy:} The privacy of customers can be exploited by analyzing their real-time electricity usage behavior; thus, information-centric networking solutions assuring data protection have to be put into practice to protect user identities.
	\item \textbf{Security:} In a fully autonomous power grid, any cyber-attack or failure in information technologies can result in billion-dollar losses; hence, security is another crucial issue.
\end{itemize}
	
As Fig.~\ref{fig:Fig1} depicts, IoEn is essential for all IoXs. Electric vehicles, as one example, impose a growing electricity demand, requiring efficient planning and management of energy from generation to utilization. The battery state and location of vehicles together with nearby charging stations' capacity/status can be tracked in real-time through the IoV to optimally match vehicles with stations, minimizing the waiting times and the pressure on scarce resources. Increasing the share of renewable sources in energy consumption can help achieve this goal while lowering greenhouse gas emissions and thus positively contribute to global Net Zero commitments. 
	
\begin{figure*}[!t]
	\centering
	\includegraphics[width=0.87\textwidth]{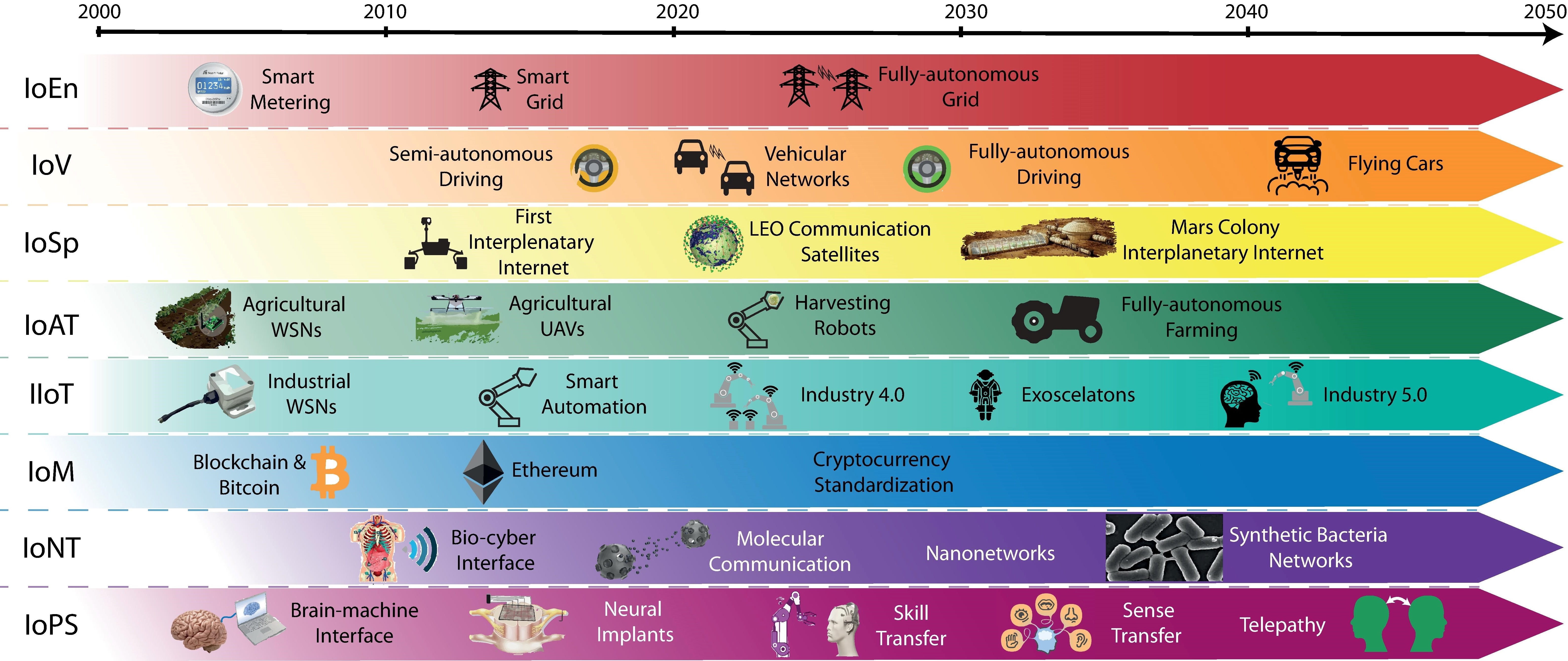}
	\caption{IoE roadmap: past, present, and future of IoE technologies.}
	\label{fig:Fig5}
\end{figure*}
	
\subsection{Internet of Space (IoSp)}
	
Communication satellites (CSs) placed in Earth's orbits~play a significant role in supporting the Internet~and are expected to be the backbone of future IoE \cite{akyildiz2019internet}. However, the existing~CS infrastructure cannot support the exploding demand for data traffic with plausible rates, and to mend this, new-age high-throughput satellites (HTSs) equipped with spot-beam technology are being launched. Besides the conventional large ($>$5tonnes) HTSs deployed in geostationary orbit (GEO), new-generation small ($<$500kg) CSs are increasingly being deployed into low earth orbit (LEO), and constellations of small CSs which network together into a dynamic web of global coverage, such as those embarked upon SpaceX and~OneWeb.
	
Recent advancements in space expedition capabilities introduced by SpaceX have made populating space with human artefacts a cheaper and faster process to the extent that a human colonization effort of Mars is no longer an unfeasible prospect. Such an effort would include CSs joining Mars Reconnaissance and Odyssey orbiters relaying information received from the Curiosity Rover with 1Gbit/day capacity, which is insufficient to support this effort. With CSs orbiting Mars deployed, the Internet on Mars can be established to join the Internet on Earth as part of a greater Internet, namely IoSp. The key challenges in IoSP realization are as below:
	
\begin{itemize}
	\item \textbf{Ubiquitous Connectivity:} Primary communication challenge in the operation of an interplanetary link involves establishing ubiquitous connectivity in a network with very high inter-node distance, which requires developing delay- and disruption-tolerant communication protocols, besides the strategic deployment of relay stations to minimize the inter-node communication disruption.
	\item \textbf{Energy Resources:} Communication devices deployed into space need to harvest energy, typically from the Sun, to perform their tasks uninterrupted. Within the IoSp framework, many communication nodes will need to be located in isolated corners of space away from the Sun, where sunlight intensity is insufficient to support communications by contemporary EH techniques. Thus, advancements in IoSp call for more efficient EH methods and low-cost communication protocols.
\end{itemize} 
	
IoSp interacts with various IoXs to improve the performance of terrestrial applications, among which the most notable ones are secure software updates for autonomous \& connected cars (w/IoV) and reliable operation of the crypto market (w/IoM), avoiding cyber-attacks and government censorship.
	
\subsection{Internet of Agricultural Things (IoAT)}
Over the last decade, we have seen the rise of smart agriculture technologies, e.g., farmland WSNs, due to the increasing food demand, food security, and health concerns. As part of the IoE framework, the IoAT covers the existing and developing information and communication technology (ICT) applications in agriculture, ranging from smart farming to smart food logistics, processing, and awareness, reinforced and diversified with the introduction of novel IoE concepts, e.g., Internet of Plants and Animals (IoPA), IoNT, IoV, that aim at the integration of every component of agriculture. 
	
Existing IoAT applications include precision agriculture with high-accuracy weather forecasts and livestock health monitoring enabled by wearable sensors, water-efficient irrigation systems, real-time tracking of individual products for food awareness and supply chain planning \cite{holden2018review}. Developing predictive analytics tools for analyzing the voluminous data generated by the heterogeneous components of the IoAT is a major topic. One of the large-scale research projects in this direction was the Internet of Food and Farm 2020, aiming to maintain safe and healthy food through IoT technologies\footnote{Internet of Food and Farm 2020. Source: \url{https://www.iof2020.eu/}}. 
	
Despite the advancements in the area, there are still major challenges hampering the realization of a full-fledged IoAT:
\begin{itemize}
	\item \textbf{Ubiquitous Connectivity and Interoperability:} IoAT requires connectivity for network components in hard-to-reach environments, e.g., remote farmlands, and seamless interoperability among heterogeneous IoAT components, e.g., intrabody IoNT nodes monitoring livestock health status, the network of plants and animals, through bio-cyber interfaces.
	\item \textbf{Big Data:} The big data generated by the IoAT components will be highly heterogeneous as it includes those generated by living entities, e.g., plants, animals, and humans, necessitating the development of novel predictive analytics tools. 
\end{itemize} 
	
Applications can be diversified with the close collaboration of IoAT with other IoE components. For example, IoNT can enable real-time and continuous health/quality monitoring of crops and livestock with molecular precision; energy-neutral drones \cite{long2018energy} (IoV) can enable automated multimedia farm-monitoring and cattle-tracking; IoSp can offer more secure, reliable, and fast farmland connectivity; IoPA, using various communication modalities, e.g., acoustic, chemical, can provide better understanding and control of the biosphere increasing the efficiency of the agricultural processes.

\section{Conclusions}
	
This paper brings the upcoming IoE revolution to attention and defines its most promising applications. There is a greater opportunity brought by the IoE vision that is still untouched because all existing IoX applications have been targeting a single domain, not requiring broad connectivity or interaction with other IoXs. To realize this opportunity, individual IoXs will be seamlessly interacting under the IoE umbrella and continuously feeding each other. In this way, we can close the technological gap between our communication infrastructure and the universe, thus fulfilling the holistic vision of the IoE.
	
The long-standing quest to develop ICT to better interact with the entire universe has enabled the large-scale deployment of WSNs over several domains starting from the early 2000s, as illustrated in Fig.~\ref{fig:Fig5}. Since we started to exploit the data collected through WSNs more effectively with advanced data analytics tools and cloud computing, we have seen the emergence of more intelligent applications, such as semi-autonomous driving and smart automation, starting from the early 2010s. Yet, this progress is not disruptive enough for developing fully automated processes covering heterogeneous technologies, e.g., fully autonomous grid, driving, farming, and manufacturing. This ambitious goal requires overcoming substantial challenges, such as universal-scale connectivity, interoperability, standardization of heterogeneous IoXs, self-sustainability, and processing of big and heterogeneous data. 
	
We have already seen some progress towards addressing the key challenges of the IoE, such as the development of brain-machine interfaces, and neural implants for interconnecting different physical domains, e.g., living cells and artificial devices; bio-compatible EH methods for self-sufficient sensors/actuators even at nanoscale; data analytics and machine learning algorithms to make use of big data for improving services. At this point, more practical steps can be taken by developing new communication techniques orthogonal to the conventional EM to extend the connectivity, devising universal transceivers that can support multiple communication modalities to provide interoperability and hybrid EH methods that can adaptively operate in different environments. Given the pace of technological advances, we are in a good position to envision a fully connected universe, including all living and artificial things, realized by 2050 via the emerging IoE~approach.

\vspace{-1.15cm}
\begin{IEEEbiographynophoto}{Ozgur B. Akan} 
received his PhD degree from the Georgia Institute of Technology, Atlanta, GA, USA, in 2004. He is currently the Head of the Internet of Everything Group, Department of Engineering, University of Cambridge, UK, and the Director of the Next-generation and Wireless Communications Laboratory, Department of Electrical and Electronics Engineering, Koc University, Turkey. His research interests include wireless, nano, molecular, and neural communications, and the Internet of Everything.
\end{IEEEbiographynophoto}
\vspace{-1.15cm}

\begin{IEEEbiographynophoto}{Ergin Dinc}
received his PhD degree in Electrical and Electronics Engineering from Koc University, Turkey, in 2016. After his PhD, he held postdoctoral positions at KTH Royal Institute of Technology, Sweden, and University of Cambridge, UK. His research interests are molecular communications, neural communication, and cyber–physical systems.
\end{IEEEbiographynophoto}
\vspace{-1.15cm}

\begin{IEEEbiographynophoto}{Murat Kuscu} 
received his PhD degrees in Engineering from University of Cambridge, UK, in 2020, and in Electrical and Electronics Engineering from Koc University, Turkey, in 2017. He is currently an Assistant Professor at the Department of Electrical and Electronics Engineering, Koc University. His research interests include Internet of Bio-Nano Things, nanomaterials, biosensors, and microfluidics.
\end{IEEEbiographynophoto}
\vspace{-1.15cm}

\begin{IEEEbiographynophoto}{Oktay Cetinkaya} 
received his PhD degree in Electrical and Electronics Engineering from Koc University, Turkey, in 2018. After his PhD, he worked as a Research Fellow at the University of Southampton, UK, a Research Associate at the University of Sheffield, UK, and a Senior Research Associate at the University of Oxford, UK. His research focuses on energy-neutral communications in the Internet of Things.
\end{IEEEbiographynophoto}
\vspace{-1.15cm}

\begin{IEEEbiographynophoto}{Bilgesu A. Bilgin} 
received his PhD degree in Mathematics from Koc University, Turkey, in 2015. After his PhD, he worked as a postdoctoral researcher at Koc University and the University of Cambridge, UK. His research interests include molecular communication, intrabody nanonetworks, and dynamical systems.
\end{IEEEbiographynophoto}

\end{document}